\documentclass{article}
\usepackage{spconf,amsmath,epsfig}

\title{Complexity Analysis of an \\Edge Preserving CNN SAR Despeckling Algorithm}

\name{Sergio Vitale $^{1}$, Giampaolo Ferraioli $^{2}$ and Vito Pascazio $^{1}$}
\address{$^{1}$ Dipartimento di Ingengeria, Universit\`{a} di Napoli Parthenope \\ $^{2}$Dipartimento di Scienze e Tecnologie, Universit\`{a} di Napoli Parthenope}

\usepackage{amsmath,graphicx}
\usepackage{subfig}
\usepackage{color}
\usepackage{epstopdf}
\usepackage{tikz}
\usepackage{balance}
\usepackage{url}
\usepackage{multirow,hhline}
\usepackage{placeins}
\usepackage{wasysym}
\usepackage{amssymb}

\usepackage[nomessages]{fp}
\usepackage{array}
\renewcommand{\aa}[1]{{\bf \textcolor{blue}{#1}}}
\newcommand{\bb}[1]{{\bf \textcolor{red}{#1}}}

\newcommand{\de}[2]{ \frac{\partial{#1}} {\partial{#2}} }

\newcommand{\labsty}[1]{{\footnotesize \textbf{#1}}}
\newcommand{\Reference}{\labsty{Reference}}
\newcommand{\Noise}{\labsty{Noisy}}

\newcommand{\labMUNOL}{\labsty{$\mathbf{M1_l}$}}
\newcommand{\labMUNOT}{\labsty{$\mathbf{M1_t}$}}
\newcommand{\labMDUEL}{\labsty{$\mathbf{M2_l}$}}
\newcommand{\labMDUET}{\labsty{$\mathbf{M2_t}$}}
\newcommand{\labMTREL}{\labsty{$\mathbf{M3_l}$}}
\newcommand{\labMTRET}{\labsty{$\mathbf{M3_t}$}}
\newcommand{\labMQUAL}{\labsty{$\mathbf{M4_l}$}}
\newcommand{\labMQUAT}{\labsty{$\mathbf{M4_t}$}}

\newcommand{\MSE}{\labsty{MSE}}
\newcommand{\SNR}{\labsty{SNR}}
\newcommand{\SSIM}{\labsty{SSIM}}
\newcommand{\Mind}{\labsty{M-index}}
\newcommand{\homo}{\labsty{H}}

\newcommand{\image}{\pgfuseimage}

\newcommand{\figpath}{./Figures/}

\definecolor{mybegie}{RGB}{128,0,0}

\pgfdeclareimage[width=0.3\textwidth]{l2}{\figpath L2.pdf}
\pgfdeclareimage[width=0.3\textwidth]{kl}{\figpath KL.pdf}
\pgfdeclareimage[width=0.3\textwidth]{edge}{\figpath EDGE.pdf}

\newcommand{\simsize}{0.15\columnwidth}

\pgfdeclareimage[width=\simsize]{sim1_noisy}{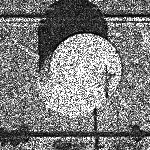}
\pgfdeclareimage[width=\simsize]{sim1_ref}{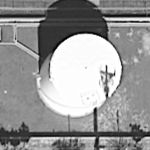}
\pgfdeclareimage[width=\simsize]{sim1_M1l}{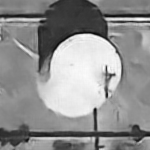}
\pgfdeclareimage[width=\simsize]{sim1_M2l}{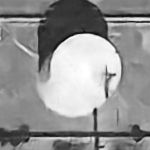}
\pgfdeclareimage[width=\simsize]{sim1_M3l}{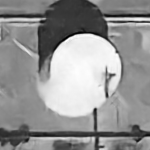}
\pgfdeclareimage[width=\simsize]{sim1_M4l}{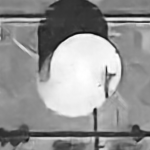}

\pgfdeclareimage[width=\simsize]{sim2_noisy}{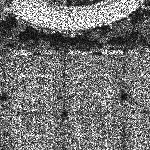}
\pgfdeclareimage[width=\simsize]{sim2_ref}{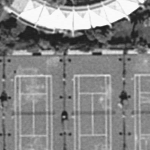}
\pgfdeclareimage[width=\simsize]{sim2_M1l}{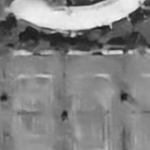}
\pgfdeclareimage[width=\simsize]{sim2_M2l}{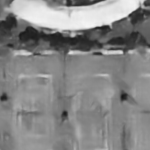}
\pgfdeclareimage[width=\simsize]{sim2_M3l}{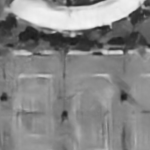}
\pgfdeclareimage[width=\simsize]{sim2_M4l}{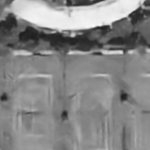}

\newcommand{\realsize}{0.18\columnwidth}

\pgfdeclareimage[width=\realsize]{real_noisy}{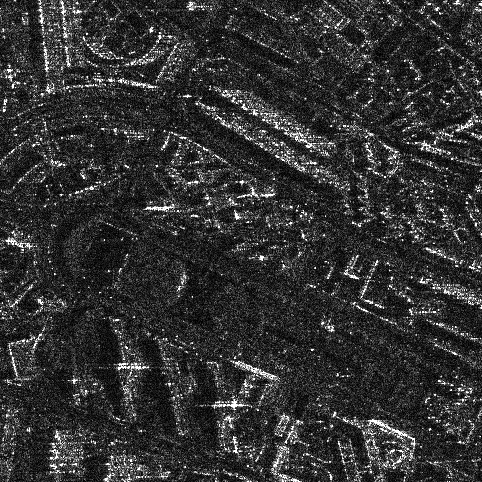}
\pgfdeclareimage[width=\realsize]{real_M1l}{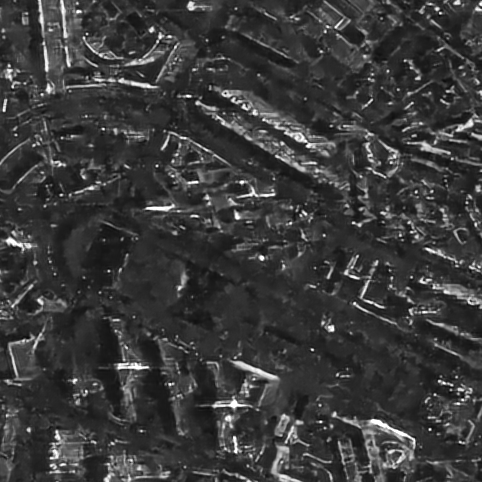}
\pgfdeclareimage[width=\realsize]{real_M2l}{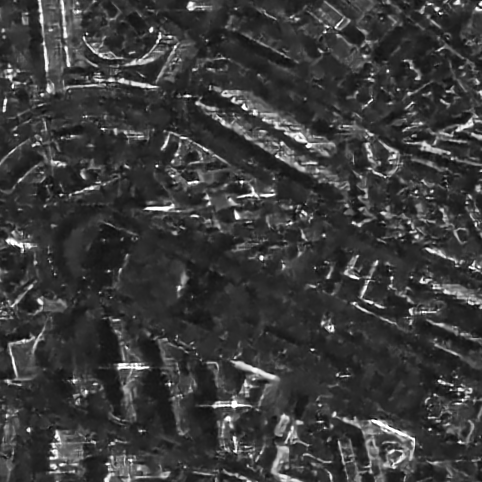}
\pgfdeclareimage[width=\realsize]{real_M3l}{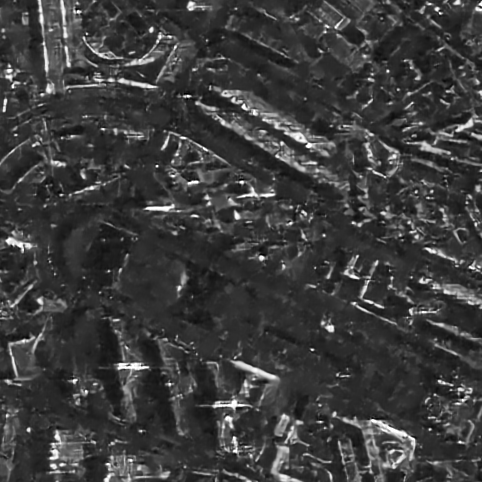}
\pgfdeclareimage[width=\realsize]{real_M4l}{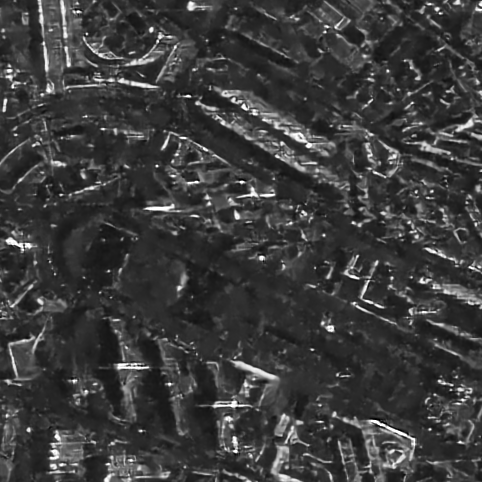}

\pgfdeclareimage[width=\realsize]{real_det_noisy}{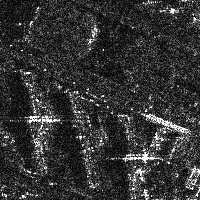}
\pgfdeclareimage[width=\realsize]{real_det_M1l}{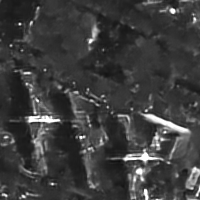}
\pgfdeclareimage[width=\realsize]{real_det_M2l}{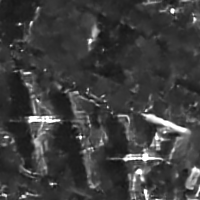}
\pgfdeclareimage[width=\realsize]{real_det_M3l}{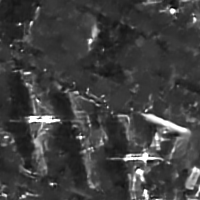}
\pgfdeclareimage[width=\realsize]{real_det_M4l}{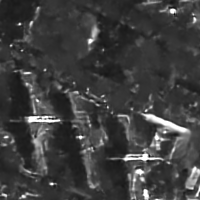}

\pgfdeclareimage[width=0.95\columnwidth]{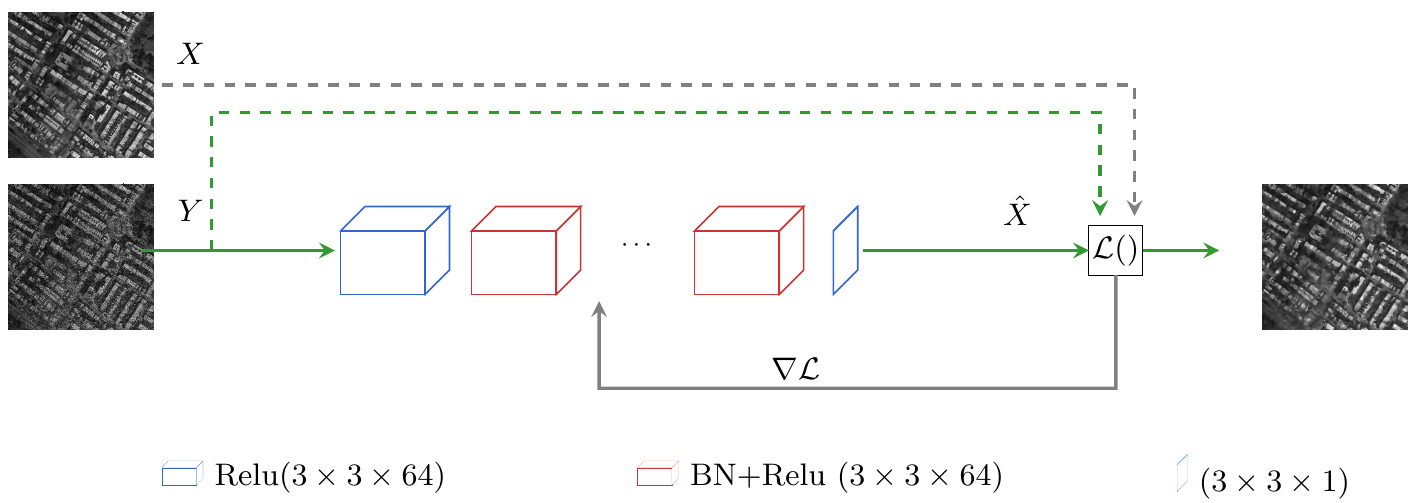}{\figpath net.pdf}

\begin{document}
%
\maketitle
\begin{abstract}
SAR images are affected by multiplicative noise that impairs their interpretations. In the last decades several methods for SAR denoising have been proposed and in the last years great attention has moved towards deep learning based solutions. Based on our last proposed convolutional neural network for SAR despeckling, here we exploit the effect of the complexity of the network. More precisely, once a dataset has been fixed, we carry out an analysis of the network performance with respect to the number of layers and numbers of features the network is composed of. Evaluation on simulated and real data are carried out. The results show that deeper networks better generalize on both simulated and real images.
\end{abstract}
\section{Introduction}
\label{sec:intro}
The several decades of investment in earth observation has made available many satellite and aerial sensors with always finer resolution and shorter revisiting time, marking remote sensing as a very important tool for remote earth investigation.

Synthetic Aperture Radar (SAR) are active sensors that acquire images continuously during day and night becoming a useful tool for several tasks such as monitoring, classification, segmentation etc.
Unfortunately, SAR images are affected by a multiplicative noise called speckle, that is due to the interference among the objects backscatterings inside a single sensor resolution cell. Hence, the resulting SAR image is composed of an alternation of bright and dark points that make its interpretation more challenging \cite{Argenti2013}.

In order to  ease further tasks such as detection, classification, segmentation and 3D reconstruction \cite{Hossi2018}, \cite{ambrosanio2017},\cite{Mazza2019}, \cite{Budillon2019}, in the last decades many studies for the implementation of despeckling filters have been conducted.
Among all the filters there is a distinction in three main categories: local, non local (NL) and CNN based filters. NL filters look for similarity among patches in all the images  according to a certain criteria \cite{Hossi2019b}. The difference between the NL filters is in the chosen similarity and combination criteria \cite{Deledalle2014}, \cite{Ferraioli2019bis},\cite{vitale2019bis}. Thanks to this rationale, NL filters ensure a better trade-off between noise reduction and edge preservation, with respect to the local filters that combine only the adjacent pixels \cite{Argenti2013}.

In the last years deep learning set a breakthrough in many fields of image processing such as classification, segmentation and detection reaching the State-of-Art performance \cite{He2017}. This improvement lured attraction in all the research fields and not less in the remote sensing community. 
Several despeckling filters based on the use of convolutional neural networks have been proposed such as \cite{Wang2017}, \cite{Chierchia2017}, \cite{Vitale2019}, \cite{Ferraioli2019},\cite{vitale2020d}. In our last proposal \cite{Vitale2019} and its improved version proposed in \cite{vitale2020}, we focused on the implementation of a new cost function. Whereas, in this work we pay attention on the network's complexity (in the sense of the amount of trainable parameters) given by the number of layers and the number of extracted features.

Obviously, the network's complexity and the dataset dimension are related each other: the aim is to exploit the performance related to the network complexity given a fixed dataset. So, we train the same architecture varying the number of layers and features and an analysis of the results have been carried out on both simulated and real images.

\section{Method}
In order to carry out the aforementioned complexity analysis, we consider as baseline the solution proposed in \cite{vitale2020}, that is a CNN of ten layers trained on simulated data. Starting from this architecture, we train several variation with different number of layers and number of features.

\subsection{Baseline}
In \cite{vitale2020}, the simulation process has been held under the fully developed hypothesis: we simulated the speckle $N$ with a Gamma distribution $p(N)$ and number of look $L=1$
$$p(N) = \frac{1}{\Gamma(L)} L^L N^{L-1} e^{-NL}$$

Once the speckle has been simulated, we multiplied it to a noise-free intensity image $X$ in order to obtain a simulated SAR image $Y = X \cdot N$

In \cite{vitale2020}, we focused mainly on the definition of a cost function that take care of spatial and spectral properties of SAR images combining three terms $$\mathcal{L} = \mathcal{L}_2 + \lambda_{edge} \mathcal{L}_{edge} + \lambda_{KL} \mathcal{L}_{KL} $$
$$ \mathcal{L}_{2} = || \hat{X} - X ||^2$$
$$\mathcal{L}_{edge} =\left( \de{X}{u} - \de{\hat{X}}{u} \right) ^2 +  \left( \de{X}{v} - \de{\hat{X}}{v} \right) ^2  $$
$$ \mathcal{L}_{KL} = D_{KL} \left( \frac{Y}{\hat{X}}, \frac{Y}{X}  \right) = D_{KL} (  \hat N, N   )  $$
where $X$,$\hat{X}$ and $\hat{N}$ are respectively the noise-free reference, the estimated noise-free image and the estimated noise. The couple $(u,v)$ indicate the horizontal and vertical direction of the image.
$\mathcal{L}_{2}$ is the MSE between the reference and the estimated noise-free image.
$\mathcal{L}_{edge}$ is a measure of the difference of the gradient along the horizontal and vertical directions of $X$ and $\hat{X}$.
$\mathcal{L}_{KL}$ is the Kullback-Leibler divergence (KLD) between the statistical distributions of the estimated noise $N$ and the theoretical one.

The first two terms are responsible of preserving spatial details and edges, respectively.
The last one involves the estimated noise in order to preserve statistical properties of the SAR image. The reader is invited to refer to \cite{vitale2020} for more details and a deeper insight.

\subsection{Proposed Analysis}

In this work we focus on the network architecture, more precisely on its depth and  width: starting from the architecture proposed in \cite{vitale2020}, we train different networks with a fixed simulated dataset.

We consider the Merced Land Use dataset \cite{MercedLandUse} and use ($ 57526 \times 64 \times 64 $) patches for training and ($14336 \times 64 \times 64$) for validation.

\begin{figure}[ht]
    \centering
    \caption{Baseline Network: it is composed of a first convolutional layer followed by ReLU, several inner convolutional layer followed by Batch  Normalization and ReLU and the output layer is a convolutional layer.}
    \image{net}
    
    \label{fig:net}
\end{figure}

In Fig.\ref{fig:net}, the basic architecture of the network is depicted. Generally, each trained network is composed of a first convolutional layer with $n_f$ extracted features followed by a ReLU as activation function. Following, there are several inner convolutional layers with $n_f$ features followed by batch normalization and by a ReLU as well. In the end, the last layer is a simple convolutional layer with a single channel output.

The kernel size ($K \times K$) of each convolutional layers is fixed to ($3 \times 3$). 
The number of layers varies from a minimum of ten to a maximum of seventeen. We were not able to train deeper networks.
For each network, we exploit the influence of number of extracted features, 
training the same architecture once with $n_f=32$ features and once again with $n_f=64$ features. Table \ref{tab: arcs} lists the trained solutions: 
\begin{itemize}
    \item $\mathbf{M\#_{t}}$ stays for thin network trained  with $n_f=32$ features
    \item $\mathbf{M\#_{l}}$ stays for large network trained  with $n_f=64$ features
    
\end{itemize}

\begin{table}[ht]
    \centering
      \setlength{\tabcolsep}{2.5pt}
      \caption{Architectures under test: network from 10 to 17 layers are considered. For each network M\#, two versions are trained: a thin version with 32 layers, named M\#$_{t}$, and a larger version with 64 features named M\#$_{l}$. In the last row the number of trainable parameters of each network expressed in base of $10^3$}. 
    \begin{tabular}{l |c|c|c|c|c|c|c|c}
        
                  & \labMUNOT & \labMUNOL & \labMDUET & \labMDUEL & \labMTRET &\labMTREL & \labMQUAT & \labMQUAL \\
         \hline
    \# layers     &10 &10 & 12 &12  &15 &15   &17 &17   \\
    \# features   &32 &64 &32 &64  &   32 &64   &32 &64\\ 
    \# parameters  &3.4 & 6.8& 4.2 & 8.3& 5.3 & 10.6 & 6.1 & 12.1 \\
    \hline
         
    \end{tabular}
    
    \label{tab: arcs}
\end{table}

\begin{figure*}[ht]
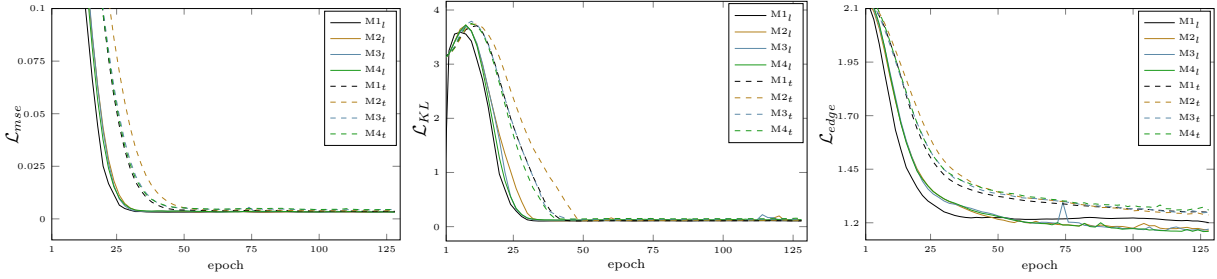

    \centering
    \caption{Loss function evaluated on validation dataset, from left to right: Mean Square Error, Kullback-Leibler Divergence, Edge Loss. Dashed line correspond to M\#$_{t}$ features network. Solid to M\#$_{l}$ features network. }
    \begin{tabular}{ccc}
          \image{l2} & \image{kl} & \image{edge}\\
    \end{tabular}
    
    \label{fig:loss}
\end{figure*}

The losses evaluated on the validation dataset are shown in Fig. \ref{fig:loss}.
According to Fig. \ref{fig:loss} the M\#$_{l}$ networks are faster and have better optimization process than the M\#$_{t}$: regarding the KLD all the network have similar performance, a slight gain is visible on the MSE and a bigger improvement is on the edge loss where M4$_{l}$ shows the best performance on both. It seems that deeper and wider network is able to better catch the properties of the image and to better filter the noise preserving spatial details.  

\section{Experiments and Results}
\begin{figure}[ht]
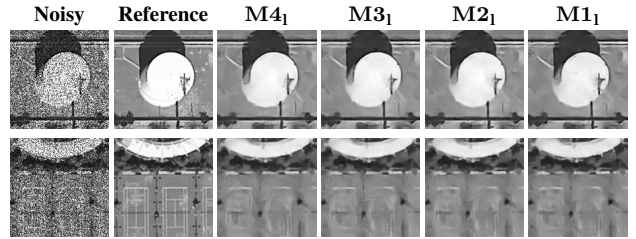

    \centering
    
    \caption{Simulated results: detail of testing images selected by Merced Land Use dataset}
    
    \begin{tabular}{cccccc}
        \Noise & \Reference & \labMQUAL & \labMTREL & \labMDUEL & \labMUNOL \\
        \image{sim1_noisy} & \image{sim1_ref} & \image{sim1_M4l} & \image{sim1_M3l} &\image{sim1_M2l}   & \image{sim1_M1l}\\
        \image{sim2_noisy} & \image{sim2_ref} & \image{sim2_M4l} & \image{sim2_M3l} &\image{sim2_M2l}   & \image{sim2_M1l}\\
    \end{tabular}
    
    \label{fig:sim-results}
\end{figure}

\begin{figure}[ht]
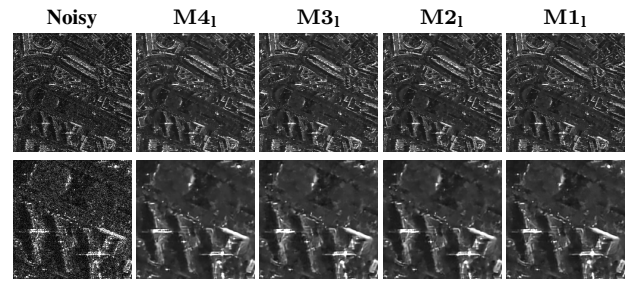

    \centering
    \caption{Real results: on the top image of Naples from CosmoSky-Med, in the bottom a detail of the full image}
    
    \begin{tabular}{ccccc}
        \Noise  & \labMQUAL & \labMTREL & \labMDUEL & \labMUNOL \\
        \image{real_noisy} &  \image{real_M4l} & \image{real_M3l} &\image{real_M2l}   & \image{real_M1l}\\
        \image{real_det_noisy} &  \image{real_det_M4l} & \image{real_det_M3l} &\image{real_det_M2l}   & \image{real_det_M1l}\\
        
    \end{tabular}
    
    \label{fig:real-results}
\end{figure}

In order to compare the performance of these solutions, we carry out experiments on both simulated and real data. Numerical and visual results are shown in Figg. \ref{fig:sim-results}-\ref{fig:real-results}. We select ($50 \times 256 \times 256$) images from the simulated dataset and one real image from COSMO-SkyMed for testing.
The networks listed in Tab.\ref{tab: arcs} are trained for 130 epochs using the Adam optimizer \cite{Kingma14} with learning rate $\eta = 0.0003$.

Tab \ref{tab:simulated results} summarizes the numerical assessment on simulated and real data. Reference metrics, such as SNR, SSIM and MSE, are averaged on the full testing dataset. For assessment on real data we consider the M-index \cite{Gomez2017} that is a combination of the ENL and Haralick homogeneity. The ENL measures the ability of suppressing noise. The homogeneity quantifies left structures in the estimated noise, so gives an indication on the ability of the filter in removing noise without suppressing useful information. In fact, pure speckle should be random and should not highlight structures, producing an homogeneity equal to 0. 
So we decide to extract the homogeneity from M-index in order to highlight such behaviour of the filters.\\
Regarding simulated evaluation, Tab\ref{tab:simulated results} shows how wider M\#$_{l}$ networks clearly outperforms the thinner M\#$_{t}$ ones on all the index, with the only M1$_{t}$ having competitive results. This behaviour is totally in line with the validation loss.
Among the wider networks M3$_{l}$ is the best one on all the index, even if all the solution are very close each other.

In order to clearly evaluate the quality of the filter, visual inspection need to accompany the numerical consideration (only the most competitive models are shown).
In Fig.\ref{fig:sim-results}, it is clear how all the M\#$_{l}$ are very close each other. It is important to notice that visually the best solution is M4$_{l}$ that shows a good edge preservation, while M3$_{l}$ and M2$_{l}$ tend to smooth results and M1$_{l}$ presents small artefacts. Indeed, M4$_{l}$ is the solution that better reconstruct the antenna on the top of the storage tank (first clip) and better try to recover the lines of tennis court.

\begin{table}[ht]
    \centering
    \setlength{\tabcolsep}{3.5pt}
    \caption{Numerical results: on the left reference metrics for simulated images (SSIM, SNR, MSE); on the right no-reference metrics for real results (M-index, homogeneity (H)); computational time for an image of size 512 $\times $512. Best value in blue, second best in red }
\begin{tabular}{l||ccc||cc||c}

                &   \SSIM & \SNR & \MSE & \Mind & \homo \tiny ($10^{-1})$ & \textbf{Time}(sec)  \\
        \hline
    \labMUNOT    & 0.7279       & 8.2276      &	294     & 9.89      & 0.293     & \aa{4}    \\
    \labMDUET    & 0.7238       & 7.9600      &	311     & 10.47     & 0.276     & \bb{4.1}  \\
    \labMTRET    & 0.7180       & 7.4201      &	352     & 10.30     & 0.218     & 4.7  \\
    \labMQUAT    & 0.7114       & 7.2222      &	366     & 8.76      & 0.117     & 4.9  \\
    \hline
    \labMUNOL    &  0.7344	    & 8.3662      & 287     & 9.30      & 0.201     & 5.2  \\
    \labMDUEL    & 0.7341       & \bb{8.5677} & 273     & \bb{8.29} & \bb{0.139}& 5.4  \\
    \labMTREL    & \aa{0.7389}  & \aa{8.6098} & \aa{270}& 9.73      & 0.247     & 5.7  \\
    \labMQUAL    & \bb{0.7375}  & 8.5635      &	\bb{272}& \aa{7.46} &\aa{0.003} & 5.8  \\

        \hline
        
\end{tabular}
    
    \label{tab:simulated results}
\end{table}

In order to have a full view of the performance of this filters, comparison on real SAR images is shown in Fig. \ref{fig:real-results}. Also in this case, the solutions present very close results with slight differences. Observing the detail in the bottom of Fig.\ref{fig:real-results}, on homogeneous area all the filters produce very similar effect. the real difference is on the treatment of the objects that produce strong backscattering. Actually, the M4$_{l}$ is the solution that better preserve objects with a slight increasing of computational time, while the other tend to smooth them making difficult the detection of single scatterers and impair the properties of adjacent areas. For comparison with other methods the reader is invited to refer to \cite{vitale2020}, where visual and numerical comparison of the M1$_l$ network with other method has been carried out.

\section{Conclusion}
In this work an analysis on the complexity of a network for SAR despeckling  has been carried out.
Starting from an our previous solution, we train different variations of the model changing  its depth (number of layers) and width (number of features).
As expected, deeper and wider trainable networks perform better.
Generally, even if there is not a big difference on the performance on simulated results, the deepest and widest network is the one that better can deal with real data. The bigger level of abstraction, the better generalization of the performance is ensured.

\label{sec:ref}

\bibliographystyle{IEEE}
\bibliography{refs}

\end{document}